\documentclass{article}

\usepackage{arxiv}
\usepackage{amsmath,amssymb,amsfonts}

\usepackage[utf8]{inputenc} 
\usepackage[T1]{fontenc}    
\usepackage{hyperref}       
\usepackage{url}            
\usepackage{booktabs}       
\usepackage{amsfonts}       
\usepackage{nicefrac}       
\usepackage{microtype}      
\usepackage{cleveref}       
\usepackage{lipsum}         
\usepackage{graphicx}
\usepackage{doi}
\usepackage{upgreek}

\usepackage{algorithmic}
\usepackage{graphicx}
\usepackage{textcomp}
\usepackage{xcolor}

\newcommand{\figref}[1]{\textsc{Figure}~\ref{#1}}
 
\usepackage{xurl}
\usepackage{comment}
\def\BibTeX{{\rm B\kern-.05em{\sc i\kern-.025em b}\kern-.08em
    T\kern-.1667em\lower.7ex\hbox{E}\kern-.125emX}}

\usepackage{tikz}
\usepackage[nolist,nohyperlinks]{acronym}

\usetikzlibrary{arrows,positioning,shapes.geometric, trees}

\title{Testbed for Functional Safety-Relevant Wireless Communication Based on IO-Link Wireless and 5G}

\author{ \href{https://orcid.org/0000-0002-6882-1214
}{\includegraphics[scale=0.06]{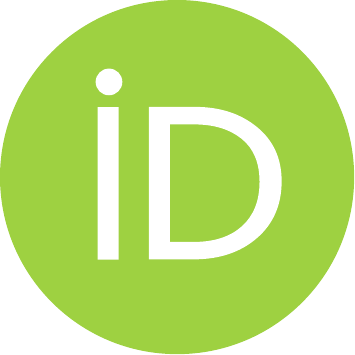}\hspace{1mm}Thomas R.~Doebbert}
\thanks{© 2022 IEEE.  Personal use of this material is permitted.  Permission from IEEE must be obtained for all other uses, in any current or future media, including reprinting/republishing this material for advertising or promotional purposes, creating new collective works, for resale or redistribution to servers or lists, or reuse of any copyrighted component of this work in other works.} \\
	Electrical Measurement Engineering\\
	Helmut-Schmidt-University\\
	Hamburg, Germany \\
	\texttt{thomas.doebbert@hsu-hh.de} \\
	\AND
    Henry Beuster \\
	Electrical Measurement Engineering\\
	Helmut-Schmidt-University\\
	Hamburg, Germany  \\
	\texttt{henry.beuster@hsu-hh.de}\\
    \And
	\href{https://orcid.org/0000-0002-4532-731X}{\includegraphics[scale=0.06]{orcid.pdf}\hspace{1mm}Florian Fischer} \\
	HSA\_innos\\
	University of Applied Sciences Augsburg\\
	Augsburg, Germany  \\
	\texttt{florian.fischer@hs-augsburg.de} \\
	\And
	Dominik Merli\\
	HSA\_innos\\
	University of Applied Sciences Augsburg\\
	Augsburg, Germany \\
	\texttt{dominik.merli@hs-augsburg.de}\\
	\And
	Gerd Scholl \\
	Electrical Measurement Engineering\\
	Helmut-Schmidt-University\\
	Hamburg, Germany  \\
	\texttt{gerd.scholl@hsu-hh.de}\\
}

\renewcommand{\shorttitle}{Testbed for Functional Safety-Relevant Wireless Communication IOLW and 5G}

\hypersetup{
pdftitle={Testbed for Functional Safety-Relevant Wireless Communication Based on IO-Link Wireless and 5G},
pdfsubject={Preprint},
pdfauthor={Thomas R. Doebbert, Henry Beuster, Florian Fischer, Gerd Scholl, Dominik Merli},
pdfkeywords={5G, IO-Link Wireless, Functional Safety, Industrial Wireless Networks, Wireless Sensor Networks},
}

\begin{document}
\maketitle

\begin{abstract}
    In the field of industrial production automation, wireless networks support highly flexible manufacturing processes and enable technologies to set-up new production chains and future software businesses. The \ac{IOLW} protocol is an already established energy-efficient and cost-effective communication standard for smart sensor devices on the industrial shop floor, whereas the mobile communication standard 5G will be mainly applied for medium and long-range wireless communication applications promising low latency times and high reliability.
    Therefore, 5G with the coming enhancement of deterministic \ac{uRLLC} is combined with the robustness and low-latency performance characteristics of \ac{IOLW}. Features of both technologies are highly beneficial to realize even highly demanding safety-related applications. The presented testbed shall qualify  wireless functional safety communication with respect to its \ac{REP} and quantify the \ac{PFH}.
    \end{abstract}
    
    
    
    \section{Introduction}
    \label{sec:Intro}
    
    
    In order to develop a suitable testbed to evaluate wireless applications in the field of functional safety, the wireless technologies employed with its specific advantages are vital to realize an all-wireless, software-defined, and safety-focused sensor-to-cloud automation system, which improves the flexibility of manufacturing processes and enhances the degree of interconnections of \acp{CPS}.
    
    Key factors such as cycle time, bandwidth, availability, reliability, deterministic communication and security of wireless channels are important regarding its applicability. Therefore, testbeds are needed to find indicators of the key performance factors. Furthermore, it is essential to develop prediction methods for reliability and latency as well as protection goals for confidentiality, integrity, authenticity, and availability of the wireless system, 
    which shall be operated in industrial environments in the domain of functional safety.
    
    Next to security enhancements on its own, also the combination of safety and security in the communication of industrial automation is a broad field for research (e.g. \cite{SchillerJuddSupavatanakulHardtWieczorek+2022+38+52, ReichenbachSafetySecurity}). In the field \ac{IOLW}, the combination of safety and security was investigated within a recent publication \cite{IOLWcrypto2022}. Other communication technologies may be investigated within future work.
    
    In the following Section \ref{sec:KeyTech}, key technologies of functional safety-relevant wireless communication are described. In Section \ref{sec:SRS}, the safety requirements with its relevant parameter and necessary calculations are evaluated and in Section \ref{sec:TestbedArchSolApproach} a possibility to validate and verify the testbed architecture solution approach is demonstrated. A conclusion and an outlook are given in Section \ref{sec:Conclusion}.
    
    \section{Key Technologies}
    \label{sec:KeyTech}
    The intended testbed with wireless safety-relevant communication, the specific technologies with features employing a testbed for safety-relevant wireless communication are presented in the following.

    \subsection{IO-Link Wireless}
    
    \ac{IOLW} is the extension of the proven IO-Link standard \cite{ILC2019} being known as \ac{SDCI} or IEC 61131-9 \cite{IEC61131}. Sensor/actuator field-bus communication within the factory automation structure is the main usage of \ac{IOLW} \cite{ILC2019,IEC61131,IOLWtestReverbChambers2022,Heynicke2018}. There are general surveys of \ac{IOLW} as an open-vendor communication solution for factory automation on the shop floor in \cite{Heynicke2018,Wolberg2018,UniHelmutSchmidt2021a} with a focus on roaming in \cite{Rentschler2017}, antenna planning in \cite{Cammin2019a}, coexistence in \cite{coexistence,IOLWcoexTool}, security enhancement in \cite{IOLWcrypto2022,IOLWcryptoPrecisionSMSI2021,iolwCryptoPerf2021}, and functional safety \cite{Doebbert2021_SafetyArchitect}, and on \ac{IOLW} testing in \cite{IOLWtestReverbChambers2022,cammin-jsss2018,jsss-8-185-2019}. Additionally, a short introduction to \ac{IOLW} is given here.
    
    \ac{IOLW} supports bidirectional wireless communication for (cyclic) process data and (acyclic) on-request data between a \ac{W-Master} and \acp{W-Device} \cite{iolw}, \cite{IEC61131} and, therefore, \ac{IOLW} is directly intended by design for fast and reliable communication on the shop-floor with dedicated technical key system properties \cite{Heynicke2018}, \cite{iolw}. \ac{IOLW} operates in the 2.4\,GHz ISM-band and its base station (i.e., \ac{W-Master}) supports a star-topology. Without performance reduction within a single manufacturing cell with a distance of up to 20\,m from the \ac{W-Master} in total up to 120 sensor or actuator nodes can operate reliably. \ac{IOLW} uses frequency hopping to mitigate fading effects due to multi-path propagation and to improve the coexistence behavior with other wireless systems \cite{coexistence}, achieving a latency below 5\,ms in typical industrial environments with a remaining failure probability of 10\textsuperscript{-9} \cite{iolw}. Therefore, the average receive power must be sufficiently high and the system must not be interfered.
    
    \subsection{5G Campus Network / Industrial 5G Campus Network}
    In the past, safety function response times in the order of 10\,ms for safety applications up to \ac{SIL}\,3 (e.g., \cite{PROFIsafeUniversity}) are critical, because response times in this range were typically not guaranteed by legacy cellular technologies such as 4G. The 5G technology shall provide better performance factors (e.g., \cite{JamesBlackmannApril2021a}) standardized in the 3GPP process \cite{Baker20200420}.
    
    The main application domains as universal communication solution for 5G networks are: \ac{eMBB} with high data rates and transmission capacities for data and streaming applications, \ac{mMTC} with a low power requirement, low complexity, low data rates and a high number and density of communication devices, and \ac{uRLLC} for communication with high reliability and very low latency \cite{EsswieDecember2020c}. In safety-relevant communication, \ac{uRLLC} is the preferred configuration being stated in Rel-16 as services regarding high levels of reliability. Rel-16 also includes the integration of \ac{TSC}, and enhancements for private network deployments being fully isolated. Private campus networks are available with Rel-15 as the first full set of 5G standards and of particular interest for industrial applications \cite{9299391,Alabbasi2019,8715451}. Further support for private networks through neutral host models is introduced in Rel-17, allowing access to standalone private networks with third-party credentials, including public network operators. Key factors are here controlability and data sovereignty \cite{5GA19} being operated e.g., with a dedicated radio access network of 3.7 GHz band width and core network functions running in premises or in the cloud respectively at the service provider.
    
    \subsection{Open Platform Communications Unified Architecture}
    \ac{OPC UA}, which is published as IEC 62541 \cite{IEC62541}, combines open-source and platform-independent industrial data-exchange with state-of-the-art IT security features for authentication, authorization and encryption. \ac{OPC UA} defines basic transport protocols for different demands and may be mapped to other Ethernet-based protocols like AMQP and MQTT \cite{OPCUAOnlineRef}. Besides the transport of live, real-time, historical or aggregated data, alerts, events and method-calls, \ac{OPC UA} provides unified information models for the multi-provider-compatible semantic description of machines, plants and entire companies. It is designed as service-oriented architecture and supports client-server as well as publish-subscribe (PubSub) communication \cite{OPCUAOnlineRef}. The core specification introduces in part 15 \cite{OPCFoundation20191031c} an IEC 61784-3 compatible \ac{SCL} to ensure deterministic data exchange between safety-related devices. Therefore, the underlaying \ac{OPC UA} channel according to the black channel-principle is used. For consistent integration into existing plants and applications several companion specification are released to unify the deployment of other communication protocols like IO-Link \cite{OPCUAIOLink} and Profinet or to standardize the information model for use-cases and markets for example Robotics and MachineVision \cite{OPCUAOnlineRef}. With regard to Industry 4.0 related topics such as cloud or edge computing and Industrial Internet of Things, the harmonized information model provides easy scalability, broad availability and high interoperability. To integrate the communication model of \ac{OPC UA} down to the shop floor, the OPC Foundation is extending the standards towards controller-to-controller and controller-to-device communication under the name of OPC UA Field eXchange (OPC UA FX), which also includes \ac{TSN}.
    
    \subsection{Time Sensitive Networking }
    A key-factor for safety-related communication is the reliable and deterministic data transmission in converged Ethernet-based networks. \ac{TSN} enables the transmission of real-time traffic besides best-effort traffic on the same Ethernet connection \cite{Jasperneite2020}.
    Next to time synchronization, bounded low latency in form of multiple schedulers, reliability and resource management are the basic parts of the \ac{TSN} standards \cite{IECJune2020c}. The specified features and key performance indicators enable convergent networks down to the field level with bump-less reconfiguration and scalable bandwidth. Due to the Ethernet-based specification, \ac{TSN} may be integrated into other communication technologies such as \ac{OPC UA} and 5G \cite{9299391} for wireless applications.
    
    This allows to converge all necessary communication streams to a single connected wire for each device.
    Multiple benefits for the domain of industrial processing are arising by this, but the development for domain specific standards e.g., IEC/IEEE 60802 for industrial automation is not completed yet.
    
    However, the trend of converged network brings also new challenges, for instance the attack surface and number of potential threats towards the real-time communication domain increases. Therefore, by using the technology of \ac{TSN} and its beneficial features, also security aspects and an effective protection strategy must be considered. Within the \ac{TSN} specification 802.1Qci is the only standard, which addresses cyber security in from of per-stream filtering and policing. This filtering is based on layer 2 (according to the ISO/OSI model), since common firewalls work on layer 3/4 or upwards and would impact the real-time capability in a non-acceptable manner. The selection and effective integration of mitigation strategies and tools is a challenging task for the \ac{TSN} domain and makes more in-depth research work necessary.
    
    \section{Measures for Safety Requirements}
    \label{sec:SRS}
    
    According to the IEC 61508 series of international standards on functional safety \cite{IEC61508}, any electrical, electronic or programmable electronic system performing safety-related functions must be developed under functional safety management to reduce the residual risks to people, equipment or the environment to an acceptable level. The required risk reduction is measured by the \ac{SIL} scaled with four steps having necessary measures and values for key factors assigned. The \ac{SIL} is evaluated in the risk analyses as the first step of the \ac{VuV} process, which describes the entire safety life cycle. Further aspects of the so-called V-model \cite{IEC61131} are, for example, the definition of the \ac{SRS}, the system, hardware and software design, test concepts for each step, verification through testing and finally the validation of the safety system of the \ac{SRS}. The safety-related system could be divided into subsystems, whereas each must meet the aspired \ac{SIL}. This enables the deployment and combination of (pre-)certified commercial hardware and software components e.g., in distributed safety functions or even networks.
    
    
    Certified fail-safe field-buses could be used to communicate between the subsystems. The corresponding IEC 61784-3 standard \cite{IEC61784-3} contains guidelines that should be followed for the exchange of safety-related data in a distributed network. Based on the black channel principle as defined in \cite{IEC61508}, no safety validation of the communication channel used is necessary when the explained techniques are applied to the \ac{SCL} and the developed \ac{FSCP}. This is advantageous for wireless connections over time-varying and frequency-selective radio channels, since there is no detailed knowledge about the channel needed. The standard recommends that the sum of failures contributed by communication should remain below one percent of the accepted \ac{PFH} of the corresponding target \ac{SIL}. Exemplary, a \ac{SIL}\,3 application with less than 10\textsuperscript{-7}\,/h dangerous failures for the entire safety-function results in a \ac{RER} for the entire \ac{SCL} of 10\textsuperscript{-9}\,/h. To comply with \cite{IEC61784-3}, communication errors related to repetition, deletion, insertion, incorrect sequence, corruption, delay, masquerade and wrong addressing must be considered. There are deterministic measures to reduce the likelihood of these communication errors such as counter/inverted counter, timeout with receipt, time expectation, communication authenticity, receipt, data integrity and e.g., redundancy with cross-check. Furthermore, models are introduced to calculate values for error rates in the domain of authenticity, timeliness, masquerade and data integrity described below. The sum of these error rates results in the total \ac{RER} ($\uplambda$\textsubscript{SC}) for the safety-channel.
    
    \subsection{Authenticity}
    
    To guarantee authenticity only correctly addressed massages from authenticated sources should be accepted. In the functional safety domain, this could be achieved by using connection-IDs as \ac{A-code} transmitted with every package. The \ac{A-code} is transmitted explicit secured by integrity measures and the fact that the rate of misrouted messages shall not exceed the message rate of the system (v), the value for the \ac{RER} for authenticity errors (RR\textsubscript{A}) may be assumed with \cite{IEC61784-3}:
    \begin{equation}
        RR_A=0 . \label{eq:RRa}
    \end{equation}
    
    \subsection{Timeliness}
    Communication errors, like delay or deletion, should be discovered to achieve the generic safety property of timeliness. Suitable methods are watchdogs, timestamps or counter values identifying the message with the \ac{T-code}. The contribution to $\uplambda$\textsubscript{SC} caused by timeliness errors is the \ac{RER} for timeliness (RR\textsubscript{T}) is
    \begin{equation}
        RR_T=2^{-LT}\cdot w \cdot R_T \cdot RP_{FCSP\_T} , \label{eq:RRt}
    \end{equation}
    with the bitlength of the \ac{T-code} (LT), the number of accepted \acp{T-code} (w), the rate of not actual messages (R\textsubscript{T}), which should be assumed in the worst case to v and additional \ac{REP} of measures regarding timeliness (RP\textsubscript{FSCP\_T}).
    
    \subsection{Masquerade}
    If a non-safety message imitates a safety message undetected, all other safety requirements for authenticity, timeliness and integrity have to be fulfilled by coincidence or accident. Thus, the \ac{RER} for masquerade (RR\textsubscript{M}) will always be low and is defined as
    \begin{equation}
        RR_M=2^{-LA} \cdot 2^{-LT} \cdot w \cdot 2^{-r} \cdot RP_u \cdot 2^{-LR} \cdot R_M , \label{eq:RRm}
    \end{equation}
    with the bitlength of the \ac{A-code} (LA), the bitlength (r) of the signature of the \ac{CRC}, the \ac{REP} for other specific data fields marking a correct safety message (RP\textsubscript{U}), bitlength of the redundant message part in case of redundancy with cross-check (LR) and the rate of occurrence of masked messages (R\textsubscript{M}), which is set to 10\textsuperscript{-3}/h for every device by default.

    \subsection{Data Integrity}
    Data integrity is the basic requirement for any safe decision; therefore, it is necessary to ensure that corruption is detected with a high and determined probability by the \ac{SCL}. In general, auxiliary redundancy must be added to double check the integrity of the data. Most popular are error-detecting codes such as \ac{CRC}, which are also proposed by the standard. For the estimation of the \ac{RER} for data integrity (RR\textsubscript{I}), the probability for an error in one bit must be assumed together with the likelihood, that this error could be detected by the selected safety measure.\\
    
    \subsubsection{Bit error probability}
    
    The black channel principle, as used according to \cite{IEC61784-3}, is based on the \ac{BSC} model.
    This model pretends, that the probability for a bit error is equal for the transmission of a digital one or digital zero at every position and could be assumed to be
    \begin{equation}
        0\leq BEP \leq0.5 . \label{eq:BEP}
    \end{equation}
    Since there would be no communication possible with a higher error probability, the standard specifies a \ac{BEP} of 10\textsuperscript{-2} to be considered unless a prove for a lower \ac{BEP} is given. Within this field of study, an ongoing discussion about the combination of safety and security measure is running in the community, which is caused by the fact that some cryptographic algorithms change the probability distribution so that the assumed \ac{BSC} is not preserved \cite{SchillerJuddSupavatanakulHardtWieczorek+2022+38+52}.\\

    \subsubsection{Properness of \ac{CRC} generator polynomials}
    The likelihood that a \ac{CRC} calculation on an erroneous message has the same result as the calculation on the original message is a degree for the properness of the \ac{CRC} generator polynomial. This probability should be calculated for every possible data length of the \ac{FSCP} explicit. If the value never exceeds the so-called conservative limit of 2\textsuperscript{-r}, where r is the \ac{CRC}-bit length, the generator polynomial is called proper and the \ac{REP} follows the calculation results.\\
    
    \subsubsection{Residual error rates}
    
    The \ac{REP} for data integrity (RP\textsubscript{I}) is given by the \ac{REP} of the proper \ac{CRC} with the specified \ac{BEP}. The equation
    \begin{equation}
        RR_I=RP_I \cdot v \cdot RP_{FSCP\_I} , \label{eq:RRi}
    \end{equation}
    with the \ac{REP} of additional safety measures for data integrity (RP\textsubscript{FSCP\_I}) is the last part to quantify the \ac{RER} for the entire safety communication layer per hour ($\uplambda$\textsubscript{SCL}) as
    \begin{equation}
        \lambda_{SCL}=(RR_T+RR_A+RR_M+RR_I) \cdot m , \label{eq:Lscl}
    \end{equation}
    with the maximum number of logical connections allowed (m) for the safety function.
    
    By applying the mentioned principles, a \ac{SRS} for a \ac{IOLS} related fail-safe \ac{IOLW} derivative is proposed and the \ac{FSCP} could be assessed using the demonstrator \cite{Doebbert2021_SafetyArchitect}.
    
    \section{Assessment of the Testbed Architecture Solution Approach}
    \label{sec:TestbedArchSolApproach}
    
    The objective of the testbed is the \ac{VuV} of the system and protocol design with the according \ac{SRS}, which is often also integrated within a safety life cycle V-model evaluation for safety-relevant measures to assess its functionality in a test environment. Further evaluation may also include safety performance and system availability \cite{9738459}. 
    
    \subsection{Validation of the Safety Requirement Specification}
    
    With the help of a certified organization, a \ac{SRS} is reviewed to survey that all risks are identified in the evaluation to meet the specification. The certified organization also reviews the \ac{SRS} regarding completeness, contradictions, and correctness. In our case, the main focus is the concept and architecture of \ac{IOLWS} rather than hardware or software implementation of the \ac{SRS}. Nevertheless, the parameter of the measures used in software to reduce risks are also reviewed. Therefore, measures such as plausibility test for the calculated $\uplambda$\textsubscript{SCL} are implemented and evaluated.
    
    \begin{figure*}[htp]
        \hspace{-1cm}
        \includegraphics[width=\textwidth]{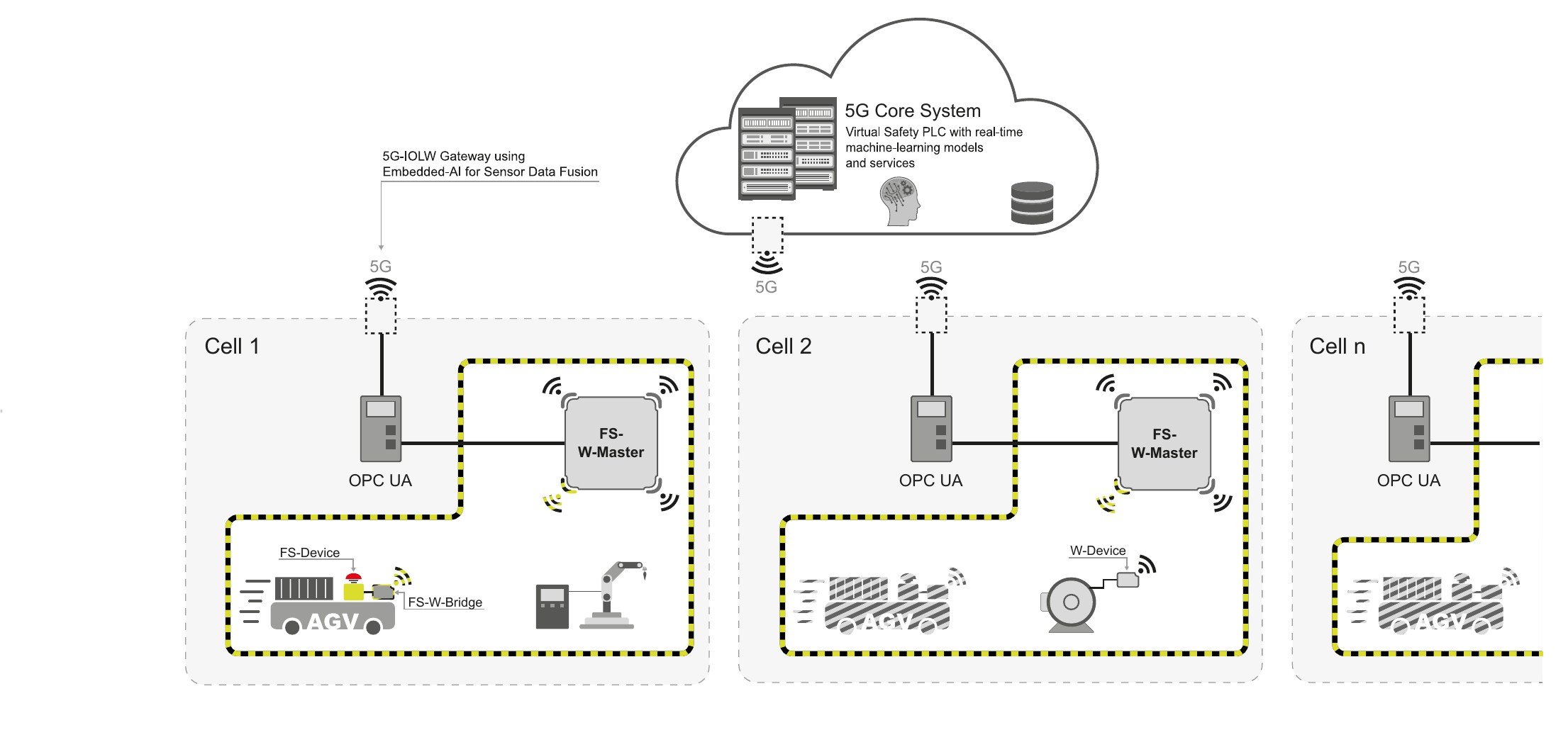}
        \caption{\textsc{Modular Sensor-2-Cloud Automation Topology (mS2CAT) for safe and secure wireless communication employing a 5G-IOLW Gateway.}}
        \label{fig:overview}
    \end{figure*}

    \subsection{Verification of System and Protocol Design}
    The verification of the testbed architecture solution approach may be realized by means of analyses, reviews, or by using a demonstrator setup, which can be based on separated lab setups. For safety functions, it is necessary to use a \ac{FDS}, which may be completed by separate function tests or complete module tests.
    
    Module tests may include hardware setup tests in a control cabinet, analysis of address ranges between modules, limit value analysis, or e.g. compliance tests with programming guidelines. In our research, the overall system design and the initial software structure is emphasizes on as well as the \ac{IOLWS} protocol design. The hardware design will not be part of the module tests.
    
    Function tests focus on the program functionality, which involve process simulations, parameter checks, and e.g. limit value tests. Therefore, various tests and analysis such as IO tests, acceptance tests, function tests, response time tests, or signal path tests are possible \cite{IEC61508}.
    
    Furthermore, \acp{FAT} for functional safety functions are possible using EN ISO 13849-2 and EN 62061.
    In our case, function, response time, and signal path tests are performed by using a demonstrator performing a specific application, which is described in the following.
    
    \subsection{Demonstrator for Functional Safety-relevant Wireless Communication}
    In \cite{9613484}, a \ac{mS2CAT} for safe and secure wireless communication employing a 5G-IOLW Gateway is described. Multiple cells are depicted for wireless communication enhancing flexible operations supporting \acp{CPS}.
    
    The \ac{mS2CAT} of \figref{fig:overview} illustrates a roaming \ac{FS-Device} being connected to an \ac{FS-W-Bridge} (in the following described as \ac{FS-W-Device}). In \cite{iolw}, the flexibility and mobility that a predefined \ac{W-Device} is able to connect to multiple predefined \ac{W-Master} cells (called roaming) is described.
    
    Depending on the location of the \ac{AGV} or person wearing a emergency stop, the \ac{FS-W-Device} is locked into Cell 1, Cell 2 or Cell n. A \ac{FS-W-Device} is only allowed to be connected to one \ac{FS-W-Master} at a time. When switching from one cell to another cell, the \ac{FS-W-Device} must first disconnect from one \ac{FS-W-Master} before connecting to another \ac{FS-W-Master}. Between the cells, a commissioning must take place to enter another safety cell depending on the application. It is not possible to enter another cell before being connected to the specific \ac{FS-W-Master}. If a safety cell is entered by accident, the cell must be set to a safe state. Time slots may vary for entering a safety cell without interrupting the manufacturing process.
    
    The \ac{FS-W-Master} in each safety cell is connected to an \ac{OPC UA} server aggregating the FS-Data using the standard master interface as \ac{FS-Gateway} according to \cite{IOLSafetySysExtensions}. The server in turn is accessed by a virtual safety \ac{PLC} integrated into the 5G core system via 5G and a 5G modem module. To set up a  continuous fail-safe communication from source to drain the mapping between \ac{IOLWS} and \ac{OPC UA} safety need to be designed compliant to \cite{IEC61784-3}, also \ac{OPC UA} PubSub and \ac{TSC} measures shall be taken into account to reduce latency and ensure deterministic response times. The described functions are combined as 5G-IOLW gateway and the \ac{OPC UA} server as software service is integrated into either the \ac{FS-W-Master} or the modem module. Both options shall be considered and compared as part of the project.
    
    The testbed with the demonstrator shall be used to test modules and functions such as the \ac{IOLWS} protocol, the response time between the 5G core system and the \ac{FS-Device} using the black channel principle through two different wireless technologies as well as in-between the protocols, the signal path within safety-relevant states and the overall functionality. Tests will be part of the future evaluation of the testbed for functional safety-relevant wireless communication. \\
    
    
    

    \section{Conclusion and Outlook}
    \label{sec:Conclusion}
    
    In this contribution, a testbed for functional safety-relevant wireless communication based on \ac{IOLW} and 5G is presented. Therefore, the well-known functional safety protocol \ac{IOLS} is used in conjunction with a new introduced wireless communication protocol \ac{IOLWS}. Furthermore, a gateway solution is enhanced to connect small-scale \ac{IOLWS} in the short-range machine-area networks with medium-scale Industrial 5G in the medium-range factory area employing technologies such as \ac{TSN} for deterministic communication and \ac{OPC UA} for safe and secure distributed communication.
    For \ac{IOLWS} the necessary \ac{SRS} is described using IEC 61508 and IEC 61784-3. After assessment of the \ac{SRS}, the testbed will be validated and verified using the described demonstrator for functional safety-relevant wireless communication.
    
    In the next step, calculations for \ac{SRS} will be evaluated and the testbed for the functional safety demonstrator will be realized.
    
    \section*{Acknowledgment}
    This work is an extended continuation of the publications under \cite{9613484, SysINT_Cammin2023}.
    The authors would like to acknowledge Christoph Cammin and Ralf Heynicke from the Helmut-Schmidt-University as well as Kunbus GmbH for their continuous support.\\
    This contribution is funded by dtec.bw – Digitalization and Technology Research Center of the Bundeswehr which we gratefully acknowledge (project "Digital Sensor-2-Cloud Campus Platform" (DS2CCP) with the project website \cite{UniHelmutSchmidt2021a}).

    \bibliographystyle{IEEEtran}
    \bibliography{Literature/literature}
    
    \begin{acronym}
        \acro{AGV}{Automated Guided Vehicle}
        \acro{IOLW}{IO-Link Wireless}
        \acro{IOLS}{IO-Link Safety}
        \acro{IOLWS}{IO-Link Wireless Safety}
        \acro{FS-Device}{IO-Link Safety Device}
        \acro{FS-W-Device}{IO-Link Wireless Safety Device}
        \acro{FS-W-Bridge}{IO-Link Wireless Safety Bridge}
        \acro{FS-W-Master}{IO-Link Wireless Safety Master}
        \acro{FS-Gateway}{Fail-Safe Gateway}
        \acro{uRLLC}{ultra-Reliable Low-Latency Communication}
        \acro{PFH}{Probability of Failure per Hour}
        \acro{CPS}{Cyber-Physical System}
        \acro{TSC}{Time Sensitive Communication}
        \acro{SDCI}{Single-Drop Digital Communication Interface}
        \acro{eMBB}{enhanced Mobile Broadband}
        \acro{mMTC}{massive Machine Type Communication}
        \acro{LTE}{Long-Term Evolution}
        \acro{OPC UA}{Open Platform Communications Unified Architecture}
        \acro{SCL}{Safety Communication Layer}
        \acro{TSN}{Time Sensitive Networking}
        \acro{SIL}{Safety Integrity Level}
        \acro{SRS}{Safety Requirement Specification}
        \acro{FSCP}{Functional Safety Communication Profile}
        \acro{RER}{Residual Error Rate}
        \acro{A-code}{Authentication-Code}
        \acro{T-code}{Timeliness-Code}
        \acro{REP}{Residual Error Probability}
        \acro{CRC}{Cyclic Redundancy Check}
        \acro{BSC}{Binary Symmetric Channel}
        \acro{BEP}{Bit Error Probability}
        \acro{FDS}{Functional Design Specification}
        \acro{FAT}{Factory Acceptance Test}
        \acro{mS2CAT}{modular Sensor-2-Cloud Automation Topology}
        \acro{PLC}{Programmable Logic Controller}
        \acro{VuV}[V\&V]{validation and verification}
        \acro{W-Master}{Wireless Master}
        \acro{W-Device}{Wireless Device}
    \end{acronym}

\end{document}